\documentclass[preprint]{revtex4-1}

\usepackage{array}
\usepackage{amsthm,amsmath}
\usepackage{graphicx}
\usepackage{hyperref}
\usepackage[utf8]{inputenc}
\usepackage{booktabs, siunitx, makecell, multirow}
\usepackage{placeins}

\begin{document}

\title{Ranking dynamics in movies and music}
\author{Hyun-Woo Lee}
\affiliation{Department of Physics, Inha University, Incheon, 22212, Republic of Korea}

\author{Gerardo I\~{n}iguez} \email[Co-corresponding author:]{gerardo.iniguez@tuni.fi}
\affiliation{Faculty of Information Technology and Communication Sciences, Tampere University, Tampere, 33720, Finland}
\affiliation{Department of Network and Data Science, Central European University, Vienna, 1100, Austria}
\affiliation{Centro de Ciencias de la Complejidad, Universidad Nacional Auton\'oma de M{\'e}xico, Ciudad de M{\'e}xico, 04510, Mexico}

\author{Hang-Hyun Jo}\email[Co-corresponding author:]{h2jo@catholic.ac.kr}
\affiliation{Department of Physics, The Catholic University of Korea, Bucheon, 14662, Republic of Korea}

\author{Hye Jin Park}\email[Corresponding author:]{hyejin.park@inha.ac.kr}
\affiliation{Department of Physics, Inha University, Incheon, 22212, Republic of Korea}

\begin{abstract} 
Ranking systems are widely used to simplify and interpret complex data across diverse domains, from economic indicators and sports scores to online content popularity. While previous studies including the Zipf's law have focused on the static, aggregated properties of ranks, in recent years researchers have begun to uncover generic features in their temporal dynamics. In this work, we introduce and study a series of system-level indices that quantify the compositional changes in ranking lists over time, and also characterize the temporal ranking trajectories of individual items' ranking dynamics. We apply our method to analyze ranking dynamics of movies from the over-the-top services, including Netflix, as well as that of music items in Spotify charts. We find that newly released movies or music items influence most the system-level compositional changes of ranking lists; the highest ranks of items are strongly correlated with their lifetimes in the lists more than their first and last ranks. Our findings offer a novel lens to understand collective ranking dynamics and provide a basis for comparing fluctuation patterns across various ordered systems.
\end{abstract}

\maketitle

\section{Introduction}
Elements of complex systems interact with each other in a complex manner, which typically leads to disparities among elements in terms of scores denoting performance, quality, or fitness of elements. Instead of studying the scores in details, one can simplify them into the ordered list, namely, by assigning ranks to elements. Such ranking might help to understand whether the performance of the system is driven by a few top ranked elements or distributed more uniformly among elements. It can also highlight statistical patterns in the orders of elements that might be obscured by fluctuations in the scores. Therefore, the representation of systems using the ranks of elements can help us understand various patterns that might otherwise remain hidden~\cite{Radicchi2009, Langville2012, merritt_scoring_2014, yucesoy_untangling_2016, Erdi2019}. Further, the increasing volume of ranking data allows us to study their common behavioral patterns at an unprecedented scale~\cite{Zipf, margalef1994, Axtell2001, adamic2002zipf, Newman01092005, Pueyo2006,  Baek_2011, Weibing2012universal, ARSHAD201875, twitchtv, Barthelemy_2023}. Examples include companies in the market, sports games, and music charts to name a few.

One of the most well-known ranking patterns is the Zipf's law~\cite{Zipf, Newman01092005}, which demonstrates that the occurrence frequency of a word is inversely proportional to its rank, and the frequency distribution follows a power law. The Zipf's law has been applied to diverse domains, such as population abundance of biological species~\cite{margalef1994, Pueyo2006}, city size~\cite{Axtell2001, ARSHAD201875}, email contact~\cite{adamic2002zipf}, and revenue of streamers~\cite{twitchtv}. Most of such works have focused on the static or time-aggregated properties of ranks.

In recent years, there has been increased attention on understanding the temporal behavior of these systems, i.e., their \textit{ranking dynamics}. One key finding is that the top ranked elements rarely change their positions in the list, while others not in the top positions change their ranks more often and to a larger extent, indicating that elements' rank changes depend on their current positions in the list~\cite{cocho2015rank,morales2018rank}. Underlying mechanisms behind different temporal patterns of ranking dynamics have also been identified~\cite{Blumm2012, temporal, Morales2018, iniguez2022}. For example, ranking dynamics observed in both open and closed systems could be successfully explained by an interplay between the movement and replacement of elements in the ranking list~\cite{iniguez2022}.

In our work, we make one step further by analyzing the daily ranking data of the movies from over-the-top (OTT) services and Spotify music charts; we focus on (i) system-level compositional changes of ranking lists and (ii) temporal trajectories of individual movie or music items. For the system-level analysis, we introduce a measure called the weighted rank change (WRC) for each item between two consecutive days, and those WRCs are aggregated to obtain the system-level rank fluctuations on the daily basis. To look at the composition of such fluctuations, we group items into three categories, i.e., influx (newly appeared items), outflux (disappeared items), and intra-flux (items staying in the list). We reveal that the influx influences most the aggregate WRC, mostly due to the newly released items having got popular immediately. For the individual-level analysis, we characterize the trajectory in terms of the lifetime, first rank, last rank, and highest rank of each item. We find that (i) the lifetime distributions are heavy-tailed, (ii) first rank distributions are more uniform than last rank distributions showing clear peaks at the bottom positions, (iii) ranks are overall positively correlated with each other, and (iv) the lifetimes tend to have the strongest positive correlation with the highest ranks. All these findings help us gain insights into the ranking dynamics of movies and music. Our analysis method can be applied to any other ranking systems in diverse backgrounds as it can reveal both common temporal patterns of ranking dynamics across datasets and unique characteristics intrinsic to each dataset.

\begin{figure}[t]
\begin{center}
\includegraphics[width=0.95\linewidth]{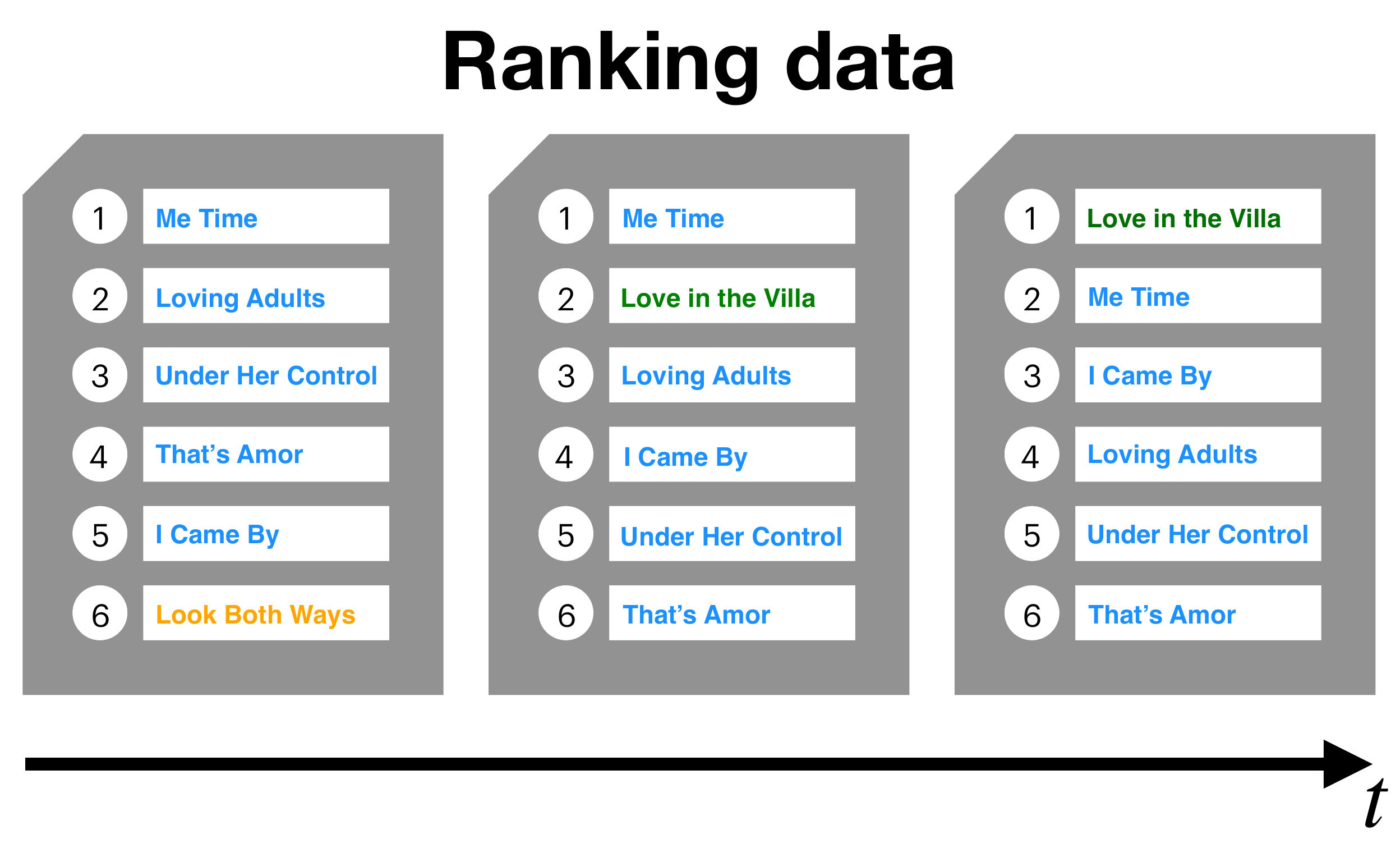}
\caption{\textbf{Dynamics of ranking.} An example of the ranking dynamics in the Netflix dataset for the first three days in September 2022, shown up to the sixth rank. A movie may disappear from the list (orange arrow), while another may enter the ranking list (green arrow). $N_{\rm b}$ denotes the number of movies in the list to be analyzed.
}
\label{fig:schematic}
\end{center}
\end{figure}

\section{Data}
To empirically study ranking dynamics of movies and music, we collect and analyze eight datasets as listed in Table~\ref{tab:info}. The first four datasets contain information on daily movie ranking from OTT services, i.e., Netflix, HBO, Disney+, and Amazon Prime. The rest four datasets are for daily music ranking from Spotify in Taiwan, UK, USA, and Australia.

\begin{table}[!ht]
    \caption{\textbf{Details of datasets.} Four daily movie ranking datasets from the OTT services, i.e., Netflix, HBO, Disney+, and Amazon Prime, and daily music ranking datasets from Spotify in Taiwan, UK, USA, and Australia. The observation period is from January 1, 2022 to December 31, 2022 for the movie data ($T = 365$ days), and from January 1, 2017 to January 9, 2018 for the music data, with four days missing ($T=370$ days). See the main text for definitions of $N_{\rm b}$ and $N_{\rm tot}$.
    }
    \label{tab:info}
    \centering
    \begin{tabular}{lcc|lcc}
       \hline
        Service  &$N_{\rm b}$ & $N_{\rm tot}$ & Service & $N_{\rm b}$ &$N_{\rm tot}$  \\ \hline
        Netflix  &83 & 2252  & Spotify Taiwan &200 & 2848 \\
        HBO &37 & 838 & Spotify UK  &200 & 2181 \\
        Disney+ &35 & 534  & Spotify USA  &108 & 975 \\
        Amazon Prime  &24 &350  & Spotify Australia  &108 & 664 \\ \hline
    \end{tabular}
\end{table}

The movie ranking data has been downloaded from the website FlixPatrol~\cite{FlixPatrol}, which provides daily worldwide movie ranking lists compiled from the top 10 movies of a number of countries, e.g., 86 countries for Netflix. In the case of Netflix, the compiled ranking list may have 860 movie items at most, but popular movies tend to appear in multiple countries at the same time, typically leading to a much smaller size of the compiled ranking list than 860. An example of ranking lists from the Netflix dataset is presented in Fig.~\ref{fig:schematic}.

We have obtained the Spotify music ranking data from Kaggle~\cite{Kaggle}. The Spotify data includes ranking lists from 45 countries. Among these, we select four countries with a substantial number of total music items, located in four different continents, i.e., Taiwan, UK, USA, and Australia.

Precisely, for each dataset, we obtain a list of items, i.e., movies or music, in a descending order on each day. Since the size of the list varies with time, we take only the first $N_{\rm b}$ items in each list of items for the consistent analysis. Here $N_{\rm b}$ is determined as the minimum of list sizes during the observation period, and it has different values for different datasets, as presented in Table~\ref{tab:info}. It means that some items whose ranks are larger than $N_{\rm b}$ are ignored for the analysis, even when they are still in the ranking list. In this sense, our analysis cannot distinguish items whose ranks are just too low from those disappeared from the list. We also obtain the total number of unique items that have ever had ranks less than or equal to $N_{\rm b}$, denoted by $N_{\rm tot}$, for each dataset in Table~\ref{tab:info}. 

Finally, we note that the determination of $N_{\rm b}$ could be seen as arbitrary; thus, we reanalyze the data using alternative values of $N_{\rm b}$ that are smaller than the original value of $N_{\rm b}$ for each dataset. We find the overall same conclusions, see Supplementary Information (SI).

\section{Methods}
\subsection{System-level and group-level analysis}
For each dataset in Table~\ref{tab:info}, we have a list of $N_{\rm b}$ items in a descending order on each day $t$, which is denoted by $L_t$, for $t=0,1,\ldots,T-1$. Here $T$ is the observation period in days, and $t=0$ indicates January 1, 2022 (January 1, 2017) for the movie data (music data). The rank of the item $i$ in the list $L_t$ is denoted by $r_{i,t}$ and it has a value among $\{1,2,\ldots, N_{\rm b}\}$. A smaller rank implies that the item is more popular. The ranking dynamics of an item can be characterized by the absolute rank change between two consecutive days as follows:
\begin{align}
    \Delta r_{i,t}\equiv |r_{i,t}-r_{i,t-1}|,
    \label{eq:rc_define}
\end{align}
where $t=1,\ldots, T-1$. From now on, the range of $t$ is $[1,T-1]$ unless otherwise stated. There can exist items that appeared in $L_t$ for the first time, thus whose ranks on the day $t-1$ are not defined. For such items, we assign their ranks to be $r_{i,t-1}=N_{\rm b}+a$ with some positive constant $a$. Similarly, items that existed in $L_{t-1}$ but disappeared on the day $t$ are assumed to have $r_{i,t}=N_{\rm b}+a$. As we have confirmed that the value of $a$ does not affect the main conclusions in our work (see SI), we use $a=1$ in the following analysis without loss of generality.

We remark that the definition of the rank change in Eq.~\eqref{eq:rc_define} cannot distinguish the impact of highly ranked items from that of lowly ranked ones when ranks of those items change. For example, if an item's rank changed from $1$ to $2$ and another item's rank changed from $100$ to $101$, their rank changes are the same as $1$ despite possibly different impacts of such changes at the system level. Thus, we introduce a weight function to attribute bigger weights to highly ranked items than to lowly ranked items:
\begin{align}
    w(r_{i,t},r_{i,t-1})\equiv \frac{1}{\min\{r_{i,t},r_{i,t-1}\}},
    \label{eq:weight_define}
\end{align}
enabling us to define the weighted rank change (WRC) as
\begin{align}
    c_{i,t} \equiv \Delta r_{i,t}\cdot w(r_{i,t},r_{i,t-1}) =  \frac{|r_{i,t}-r_{i,t-1}|}{\min\{r_{i,t},r_{i,t-1}\}}.
    \label{eq:wrc_define}
\end{align}
Returning to the mentioned example, if the item's rank changed from 1 to 2, it gives $c_{i,t}=1$, while if the rank changed from 100 to 101, we get $c_{i,t}=1/100$. As for the weight function in Eq.~\eqref{eq:weight_define}, one can instead use other functional forms such as $w(r,r')=1/\sqrt{rr'}$ or $(r+r')/(rr')$~\cite{Murase2019Sampling}, which are found to give the qualitatively similar results (see SI).

The individual WRC can be aggregated to characterize the macroscopic ranking dynamics of the system. We define the aggregate WRC as
\begin{align}
    c_t \equiv \sum_{i\in L_t\cup L_{t-1}} c_{i,t},
    \label{eq:ct_define}
\end{align}
where $L_t$ denotes the list of items on the day $t$. The aggregate WRC is expected to quantify the overall fluctuation of the ranking dynamics at the system level.

To bridge the gap between the individual WRC and the aggregate WRC, we group items for each day into three categories, namely, influx, outflux, and intra-flux. The influx items refer to those that did not exist on the day $t-1$ but appeared on the day $t$. The outflux items refer to those that existed on the day $t-1$ but disappeared on the day $t$. Finally, the intra-flux items are those that existed both in $L_t$ and $L_{t-1}$. These fluxes are respectively defined as
\begin{align}
    c^{\rm in}_t \equiv \sum_{i\in L_t\setminus L_{t-1}} c_{i,t},\ c^{\rm out}_t \equiv \sum_{i\in L_{t-1}\setminus L_t} c_{i,t},\ c^{\rm intra}_t \equiv \sum_{i\in L_t\cap L_{t-1}} c_{i,t}.
    \label{eq:cinoutintra_t_define}
\end{align}
Then it is straightforward to show that the aggregate WRC in Eq.~\eqref{eq:ct_define} is simply the sum of influx, outflux, and intra-flux WRCs:
\begin{align}
    c_t = c^{\rm in}_t + c^{\rm out}_t + c^{\rm intra}_t.
    \label{eq:c_compose}
\end{align}
An example of the influx (outflux) item in the Netflix dataset is the movie ``Love in the Villa'' (``Look Both Ways'') as shown in Fig.~\ref{fig:schematic}. All other movies in Fig.~\ref{fig:schematic} are intra-flux items. Note that ``Love in the Villa'' is the influx item on September 2nd, while it becomes the intra-flux item on the next day. This categorization turns out to be useful in understanding, e.g., which flux contributes most to the overall behavior of $c_t$.

Then we measure the fraction of each flux to the aggregate WRC for the entire range of period as follows:
\begin{align}
    g^{\alpha}\equiv \frac{\sum_{t=1}^{T-1}c^{\alpha}_t}{\sum_{t=1}^{T-1} c_t},
    \label{eq:all_fract_define}
\end{align}
where $\alpha\in\{\rm in, out, intra\}$. Note that $g^{\rm in}+g^{\rm out}+g^{\rm intra}=1$. To look at the contribution of each flux to the large aggregate WRC, we sort $c_t$ for $t=1,\ldots,T-1$ in a descending order and then cluster them into ten groups of the roughly same size of $T/10$. The groups are denoted by $G_j$ for $j=1,\ldots,10$; the group of $j=1$ corresponds to the top 10\% of $c_t$s, while the group of $j=10$ does to the bottom 10\% of $c_t$s. For each group indexed by $j$, we calculate the fractions of influx, outflux, and intra-flux WRCs to the aggregate WRC as follows:
\begin{align}
    g^{\alpha}_j\equiv \frac{\sum_{t\in G_j}c^{\alpha}_t}{\sum_{t\in G_j} c_t},
    \label{eq:fract_define}
\end{align}
where $\alpha\in\{\rm in, out, intra\}$. Note that $g^{\rm in}_j+g^{\rm out}_j+g^{\rm intra}_j=1$ for all $j$s.

Finally, to detect the periodic patterns in the time series of the aggregate WRC, we calculate the power spectrum density $P(f)$ of the time series of $c_t$, as follows~\cite{stoica2005spectral}: 
\begin{align}
    P(f)\equiv \frac{2}{T-1} \left| \sum_{t=1}^{T-1} c_t e^{-2\pi if(t-1)} \right|^2.
        \label{eq:power_define}
\end{align}

\subsection{Individual-level analysis}
We also analyze the ranking dynamics data at the level of individual items. For each item in each dataset, we obtain its lifetime, first rank, last rank, and highest rank. Since the lifetimes of items appeared on the first day ($t=0$) and/or the last day ($t=T-1$) of the observation period might be censored, we exclude those items from the analysis. 
Let $t^{\rm first}_i$ ($t^{\rm last}_i$) denote the date when the item $i$ appeared on the ranking list for the first (last) time. Then the lifetime $l_i$ is defined as
\begin{align}
    l_i\equiv t^{\rm last}_i-t^{\rm first}_i.
    \label{eq:lifetime}
\end{align}
The first (last) rank of the item $i$ is the rank on the day $t^{\rm first}_i$ ($t^{\rm last}_i$), while the highest rank of the item $i$ means the highest rank achieved during its lifetime, namely,
\begin{align}
    r^{\rm first}_i\equiv r_{i,t^{\rm first}_i},\ r^{\rm last}_i\equiv r_{i,t^{\rm last}_i},\ r^{\rm high}_i\equiv \min_{t^{\rm first}_i\leq t\leq t^{\rm last}_i} r_{i,t}.
    \label{eq:first_highest_rank}
\end{align}
By definition, $r^{\rm high}_i\leq r^{\rm first}_i$ and $r^{\rm high}_i\leq r^{\rm last}_i$ for each $i$. We shall study the distributions of these quantities as well as the correlation patterns between them.

\section{Results and Discussion}

\subsection{System-level and group-level analysis}

\begin{figure}[!ht]
\begin{center}
\includegraphics[width=0.95\linewidth]{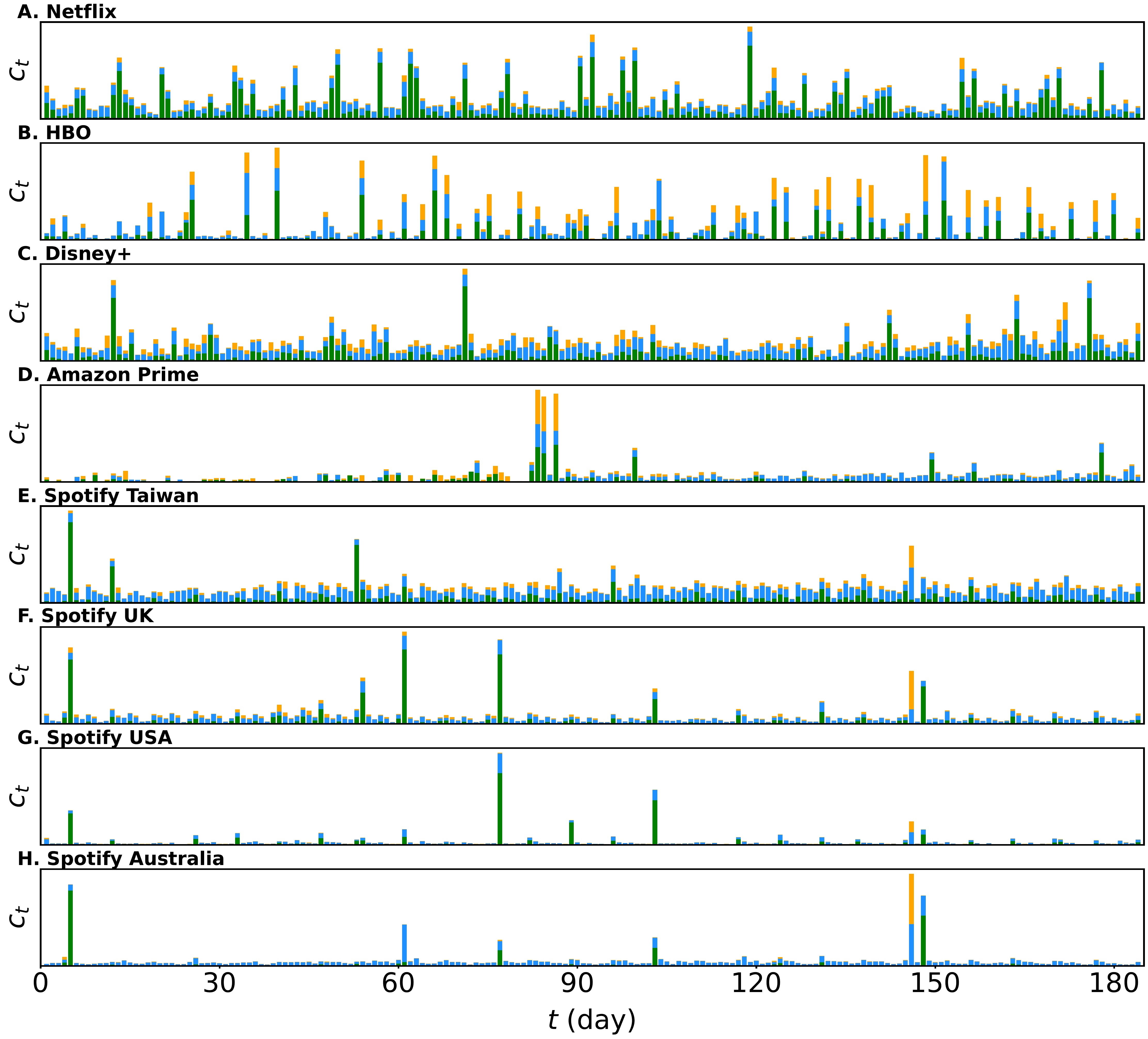}
\caption{\textbf{Time series of the aggregate WRC.}
Daily time series of the aggregate weighted rank change (WRC) $c_t$ in Eq.~\eqref{eq:ct_define} for the first six months derived from eight datasets in Table~\ref{tab:info}. $t=0$ indicates the first day of the observation period in each dataset. Each bar for $c_t$ is decomposed into influx (green), outflux (orange), and intra-flux (blue) WRCs in Eq.~\eqref{eq:cinoutintra_t_define}. 
}
\label{fig:WRC}
\end{center}
\end{figure}

\begin{figure}[!ht]
\begin{center}
\includegraphics[width=0.95\linewidth]{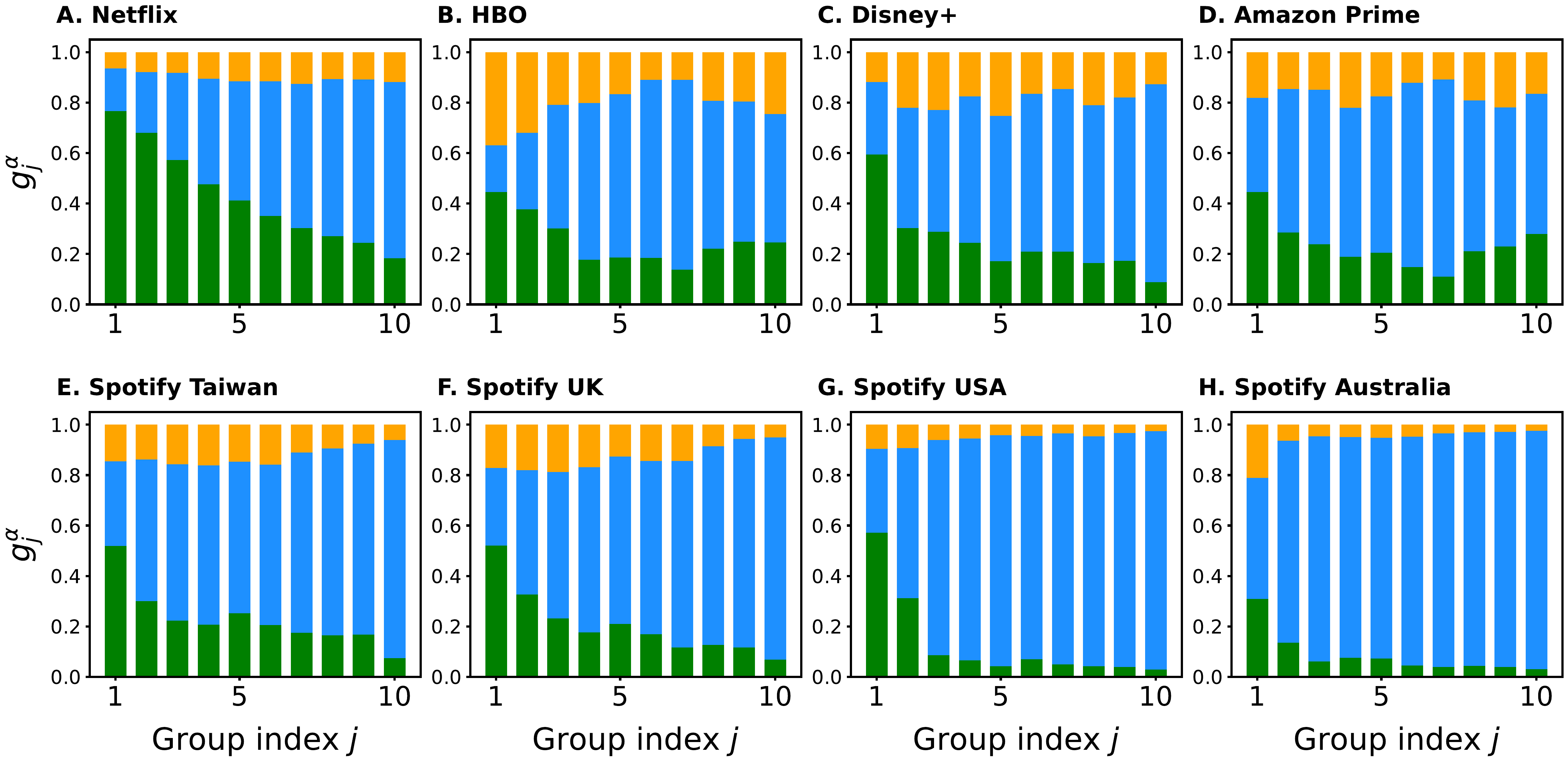}
\caption{\textbf{Fraction of fluxes.}
Fractions of influx (green), outflux (orange), and intra-flux (blue) WRCs to the aggregate WRC for each of ten groups, i.e., $g^{\alpha}_j$ for $\alpha\in\{\rm in, out, intra\}$ and $j=1,\ldots,10$ in Eq.~\eqref{eq:fract_define}. The group of $j=1$ ($j=10$) indicates the top (bottom) 10\% of all $c_t$ values during the observation period.
}
\label{fig:group_fraction}
\end{center}
\end{figure}

For each dataset, we calculate the aggregate weighted rank change (WRC) $c_t$ in Eq.~\eqref{eq:ct_define} for the entire observation period. We plot the time series of $c_t$ for the first six months in each dataset in Fig.~\ref{fig:WRC} due to the limited space. According to Eq.~\eqref{eq:c_compose}, each bar of $c_t$ in Fig.~\ref{fig:WRC} is decomposed into three fluxes, each of which is colored either green (influx), orange (outflux), or blue (intra-flux). We find various temporal fluctuation patterns of $c_t$; the time series of $c_t$ for Netflix, HBO, and Spotify Taiwan show overall periodic behaviors, while other datasets tend to show intermittent peaks amid noisy backgrounds. Such periodicity has been confirmed by the power spectrum analysis in Eq.~\eqref{eq:power_define}, to be discussed later in more detail.

The composition of influx, outflux, and intra-flux WRCs for the aggregate WRCs also shows different patterns. We first calculate the fractions of three fluxes to the aggregate WRC for the entire range of period in Eq.~\eqref{eq:all_fract_define}, as summarized in Table~\ref{tab:contribution}. We observe for all datasets that $g^{\rm in}$ is greater than $g^{\rm out}$, implying that the influx items play a more important role than outflux ones in the ranking dynamics at the system level. Then, we calculate the fractions of three fluxes in the aggregate WRCs for 10 groups of $c_t$s as defined in Eq.~\eqref{eq:fract_define}. The results are shown in Fig.~\ref{fig:group_fraction}. In all datasets except for the Spotify Australia, the influx WRCs contribute the most to the top 10\% of $c_t$ as $g^{\rm in}_1$ ranges for 0.45$-$0.75; in the case with the Spotify Australia, we find $g^{\rm in}_1\approx 0.3$. It implies that peaks found in the time series of $c_t$ are mostly attributed to the influx items, in particular, when those items occupy the top ranks as soon as they appear in the list. We also observe in Fig.~\ref{fig:group_fraction} that the fraction of the influx WRC, i.e., $g^{\rm in}_j$, is overall decreasing for the increasing $j$, implying that influx items contribute less and less for the groups of low $c_t$s.

\begin{table}[!ht]
    \centering
    \caption{\textbf{Contributions of fluxes to aggregate WRC.}
    Contributions of influx, outflux, and intra-flux WRCs to the aggregate WRC for the entire observation period in Eq.~\eqref{eq:all_fract_define} for each dataset. The numbers are rounded.}
    \begin{tabular}{lccc|lccc}
        \hline
        Service &$g^{\rm in}$  &$g^{\rm out}$ & $g^{\rm intra}$ & Service 
        & $g^{\rm in}$  &$g^{\rm out}$ & $g^{\rm intra}$\\ \hline   
        Netflix &0.56  &0.09 & 0.35 & Spotify Taiwan & 0.30 & 0.14 & 0.56 \\
        HBO &0.36 &0.30 & 0.34  & Spotify UK &0.33 &0.16 & 0.51\\
        Disney+ &0.32  &0.18 & 0.50 & Spotify USA &0.37 &0.08 & 0.55\\
        Amazon Prime &0.30 &0.17 &0.53 & Spotify Australia &0.18 &0.12 &0.70 \\ \hline
        \end{tabular}
    \label{tab:contribution}
\end{table}

\begin{figure}[!ht]
\begin{center}
\includegraphics[width=0.95\linewidth]{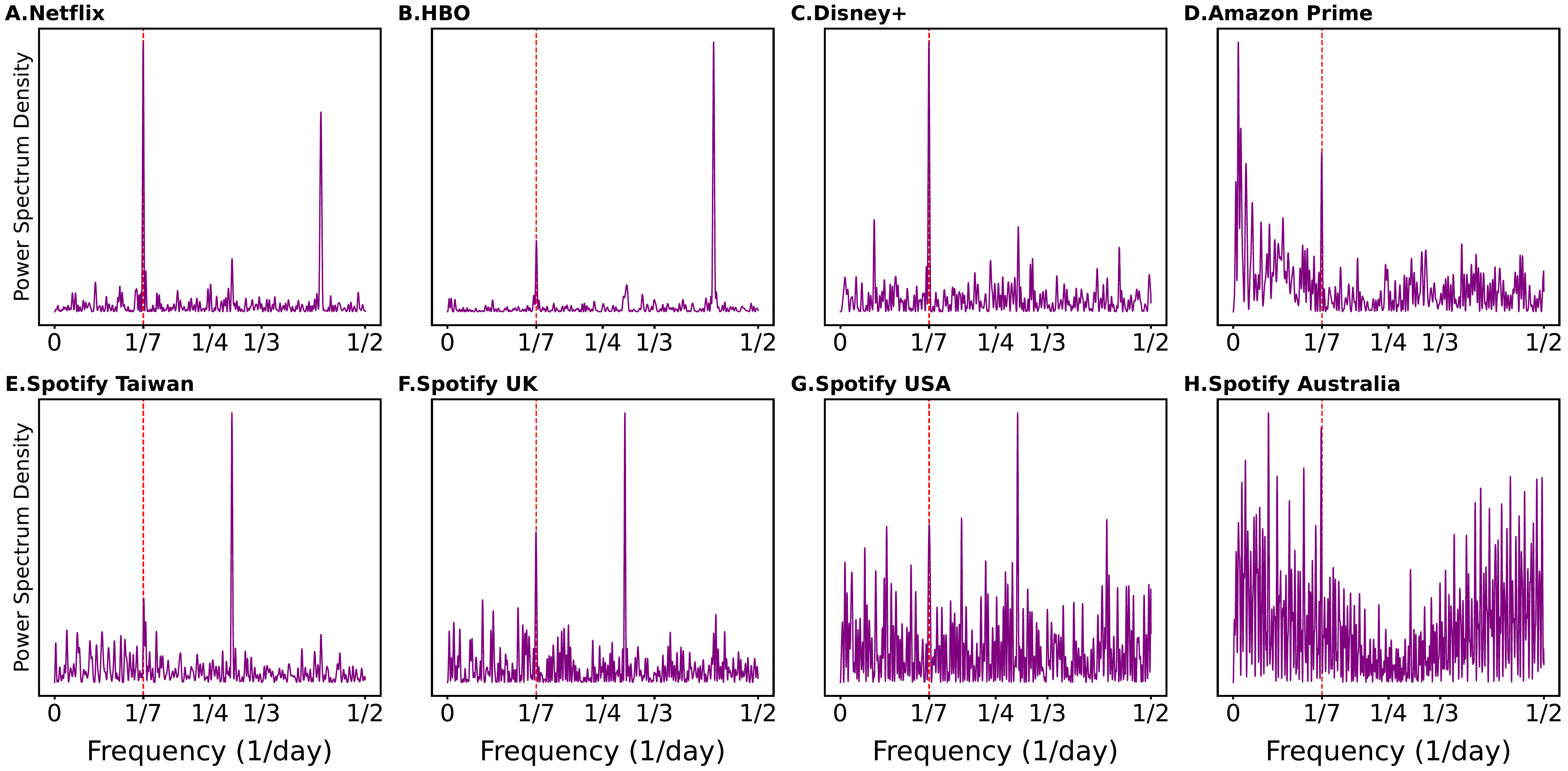}
\caption{\textbf{Power spectrum densities of the aggregate WRC $c_t$ for each dataset.}
Red vertical dashed lines are plotted at the frequency corresponding to a week for guiding eyes.
}
\label{fig:power}
\end{center}
\end{figure}

Next, we examine the periodic patterns of the aggregate WRC $c_t$ by calculating their power spectrum density $P(f)$ in Eq.~\eqref{eq:power_define}. The results are shown in Fig.~\ref{fig:power}, where we find peaks at the frequency corresponding to a week in all datasets, whether they are the highest or the second to the highest. We also find other peaks appeared at different frequencies depending on the dataset. Precisely, Netflix and HBO data show peaks corresponding to 2--3 days. Amazon Prime data shows the peak corresponding to $\sim$365 days, which however seems artifactual considering $T=365$ days. In contrast, all Spotify datasets show peaks of 3--4 days, except for Spotify Australia with a less distinct peak of 3--4 days.

\begin{figure}[!ht]
\begin{center}
\includegraphics[width=0.95\linewidth]{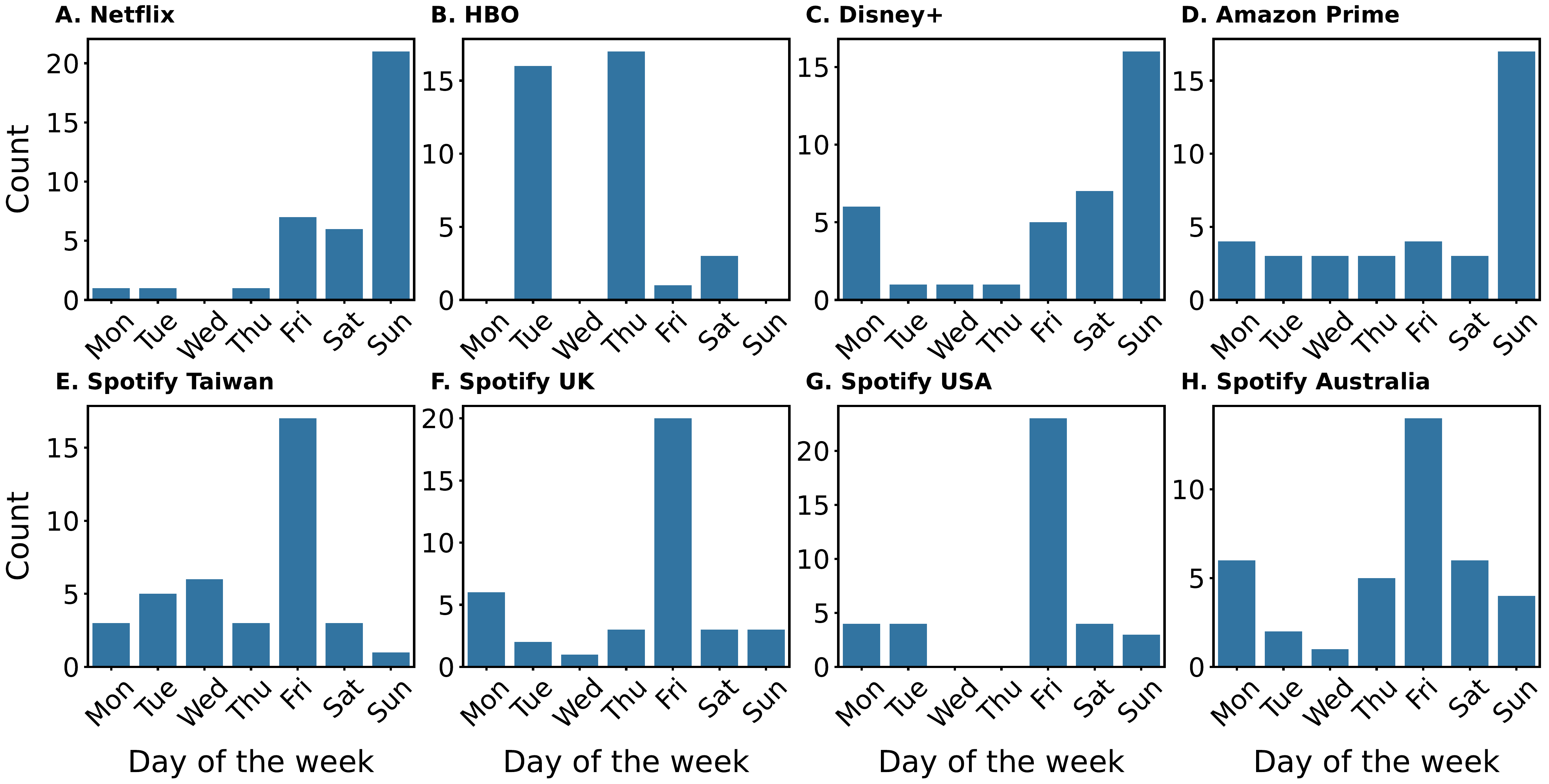}
\caption{\textbf{Distributions of days of the week.}
Distributions of days of the week of $t$s for the top 10\% of all $c_t$ values during the observation period, i.e., for $\{t|c_t\in G_1\}$.
}
\label{fig:days}
\end{center}
\end{figure}

We briefly discuss about the possible reasons behind the periodic behaviors of the aggregated WRC observed in Fig.~\ref{fig:power}. In cases of movies, the observed weekly periodicity may be influenced not only by the nature of movie industry but also by the service users' behavioral patterns~\cite{MaC9362}. Users possibly tend to watch movies more on weekends than on weekdays, leading to the major changes of movie charts. To test this idea, we measure the distributions of days of the week of $t$s for the top 10\% of all $c_t$ values during the observation period, i.e., for $\{t|c_t\in G_1\}$, as shown in Fig.~\ref{fig:days}A--D. We indeed find that in the Netflix, Disney+, and Amazon Prime datasets, major changes of movie charts occur mostly during weekends, while for HBO, such changes occur mostly on Tuesdays and Thursdays, being consistent with the dominant peak at 2--3 days of the power spectrum density in Fig.~\ref{fig:power}B. The distributions of days of the week for the Spotify datasets show dominant peaks on Fridays as well as other smaller peaks on Mondays or Wednesdays in Fig.~\ref{fig:days}E--H. This might be due to spatiotemporal constraints for consuming music to be more relaxed than for watching movies~\cite{6566767}.

\begin{figure}[!ht]
\begin{center}
\includegraphics[width=0.95\columnwidth]{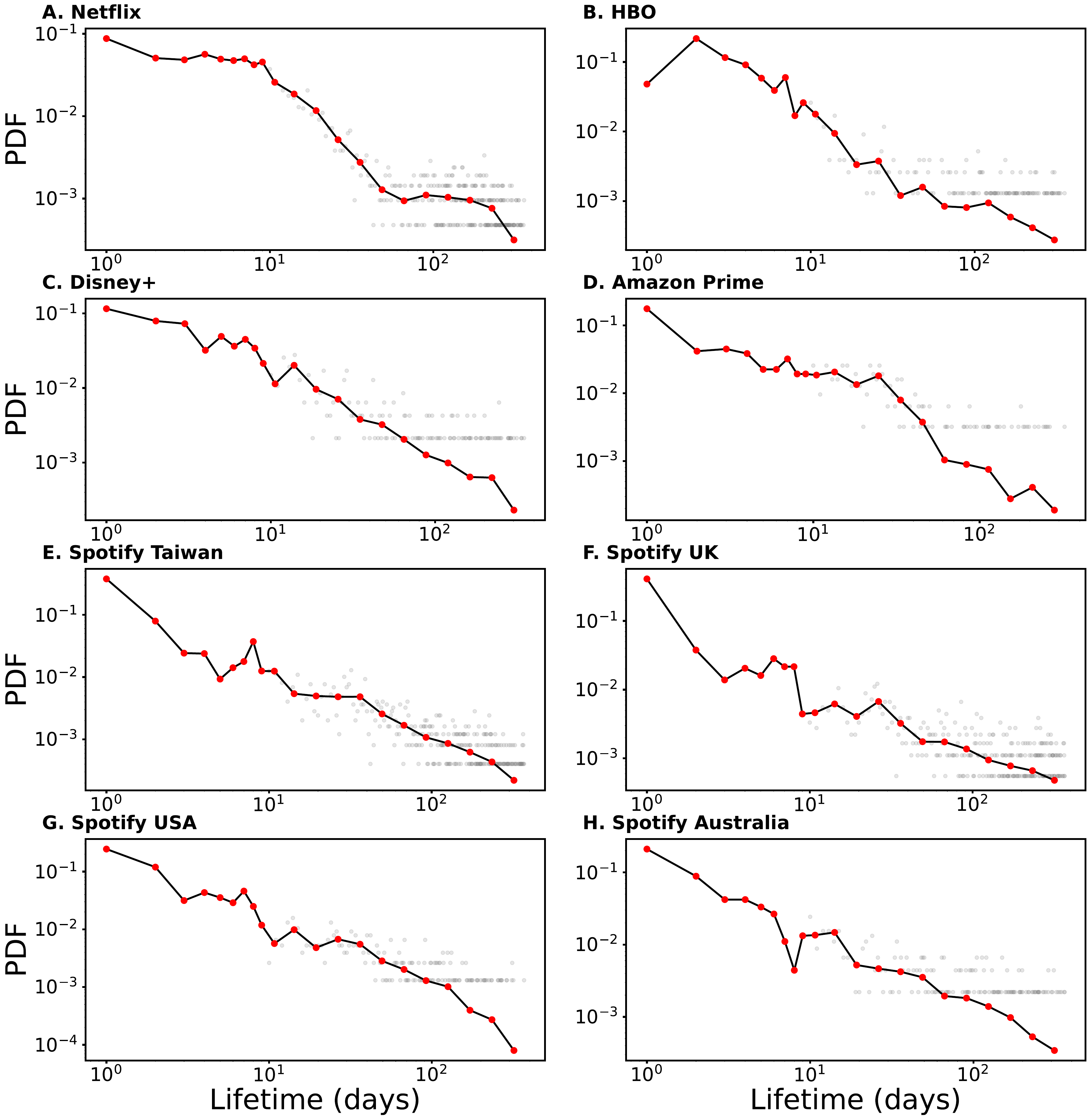}
\caption{\textbf{Distributions of lifetimes.}
Gray dots and red filled circles denote the original and log-binned distributions, respectively.
}
\label{fig:life}
\end{center}
\end{figure}

\begin{figure}[!ht]
\begin{center}
\includegraphics[width=0.9\linewidth]{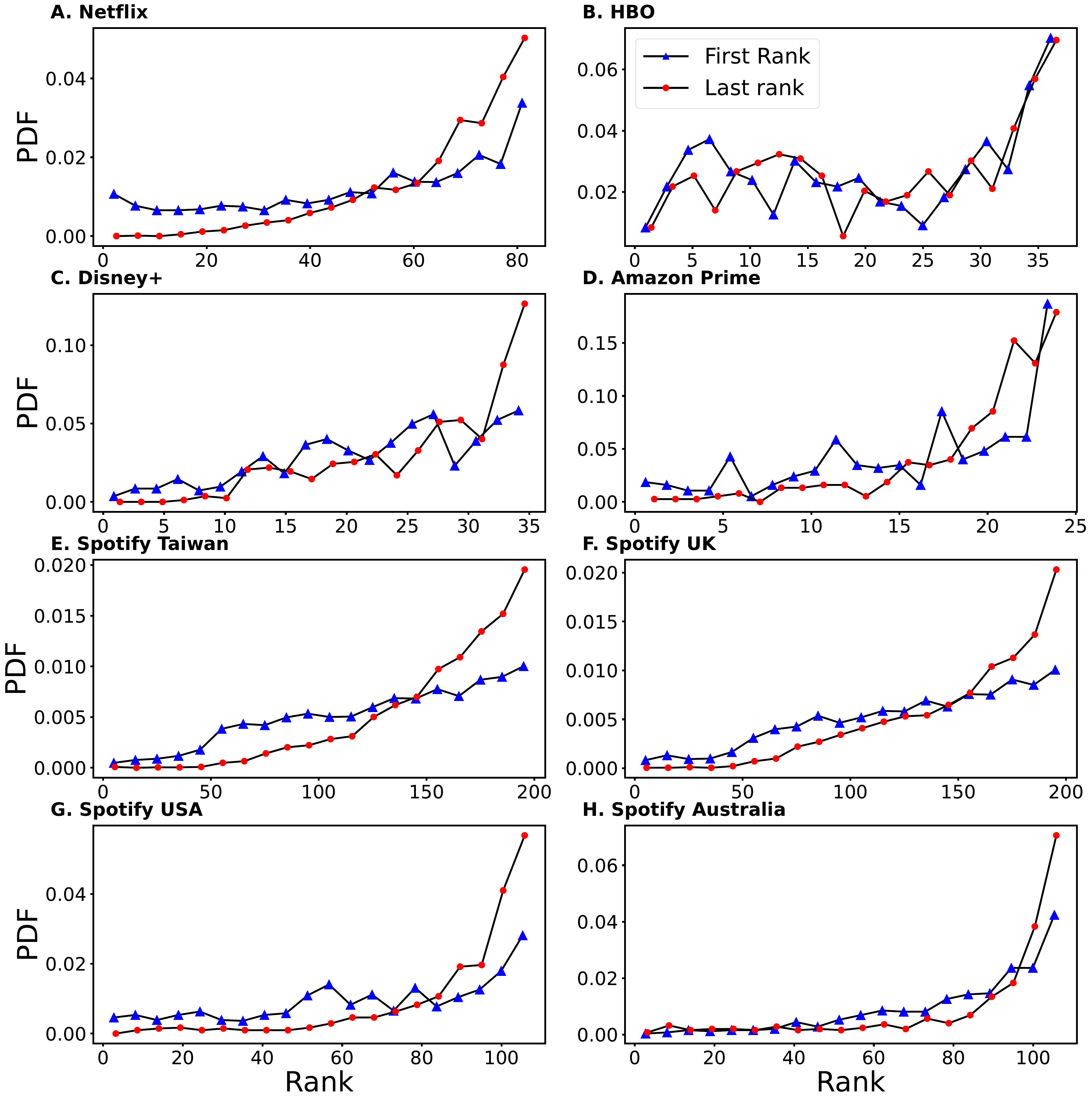}
\caption{\textbf{Distributions of first and last ranks.}
Blue triangles and red circles denote the distributions of first and last ranks, respectively. Solid lines are provided as visual guides.
}
\label{fig:first-last}
\end{center}
\end{figure}

\subsection{Individual-level analysis}
We study ranking dynamics at the individual level by looking at each item's lifetime, first rank, last rank, and highest rank, as defined in Eqs.~\eqref{eq:lifetime} and~\eqref{eq:first_highest_rank}.

We first obtain the lifetime distributions from datasets analyzed. As shown in Fig.~\ref{fig:life}, all distributions show heavy tails over the broad range of lifetimes, implying the presence of items staying in the ranking lists for a much longer time than others. These heavy tails might be understood not only by items' intrinsic properties but also by the competition between items~\cite{Wu2007, crane2008}.

Next, we calculate the distributions of first and last ranks as depicted in Fig.~\ref{fig:first-last}. All distributions consistently show increasing trends in ranks, implying that items typically enter and exit the ranking list at lower positions. This observation aligns with the previous finding that ranks tend to be more stable at the top than at the bottom~\cite{iniguez2022}. Furthermore, first rank distributions are overall more uniform than last rank distributions having clear peaks at the bottom positions for all datasets. It implies that items tend to appear in the ranking list more uniformly but disappear mostly at the bottom, which characterizes the dynamics of chart turnover driven by the release of new items. Precise correlation patterns between first and last ranks of the same items will be discussed below.

\begin{table}[!ht]
\centering
\caption{\textbf{Pearson correlation coefficients.} 
    Pearson correlation coefficients between lifetimes, first ranks, last ranks, and highest ranks of items for each dataset.}
\begin{tabular}{lcccccc}
\toprule\hline
{Service} & {life–highest} & {life–first} & {life–last} & {first–highest} & {first–last} & {last–highest} \\
\midrule\hline
Netflix     & -0.28 & -0.02 &  0.02 &  0.73 &  0.27 &  0.39 \\
HBO        & -0.29 &  0.02 &  0.08 &  0.77 &  0.68 &  0.70 \\
Disney+      & -0.44 & -0.18 & -0.03 &  0.76 &  0.35 &  0.46 \\
Amazon Prime       & -0.38 & -0.17 &  0.10 &  0.74 &  0.46 &  0.34 \\
Spotify Taiwan & -0.42 & -0.18 &  0.14 &  0.77 &  0.26 &  0.19 \\
Spotify UK & -0.37 & -0.16 &  0.08 &  0.74 &  0.05 &  0.18 \\
Spotify USA & -0.40 & -0.20 &  0.18 &  0.73 & -0.07 &  0.15 \\
Spotify Australia & -0.52 & -0.25 &  0.25 &  0.60 & -0.01 &  0.25 \\
\bottomrule\hline
\end{tabular}
\label{tab:pcc}
\end{table}

As for the correlation patterns between lifetimes, first ranks, last ranks, and highest ranks, we measure the Pearson correlation coefficients (PCCs) between all possible pairs of those quantities. The results are summarized in Table~\ref{tab:pcc}. Overall, first, last, and highest ranks are positively correlated with each other in most cases, while slightly negative correlations are also observed between first and last ranks for the Spotify USA and Australia. To be precise, the PCCs between first and highest ranks are consistently high, i.e., larger than 0.7 across datasets, except for Spotify Australia. One can expect such strong correlations by the inequality between those ranks, namely, $r^{\rm high}_i\leq r^{\rm first}_i$. However, this is not the case with the PCCs between last and highest ranks, whose values considerably vary from 0.15 to 0.7 depending on the datasets despite the fact that $r^{\rm high}_i\leq r^{\rm last}_i$. The PCCs between first and last ranks also show a wide range of values.

We then observe that lifetimes are most strongly correlated with the highest ranks; more popular items tend to stay longer in the list. The PCCs between lifetimes and first (last) ranks are overall negative (positive), except for HBO (Disney+), but their magnitudes of correlation are consistently smaller than those between lifetimes and highest ranks. This indicates that the highest rank could be a better predictor of item longevity than both first and last ranks. 

We finally note that the last ranks show overall weak correlations with other quantities compared to those between lifetimes, highest ranks, and first ranks, and those PCC values show most inconsistent signs and magnitudes. It implies that the last ranks may not be a good predictor for other ranks and lifetimes of items.

\section{Conclusion}
We have analyzed eight datasets for ranking dynamics of movies and music items both at the system level and at the individual level. For the system-level analysis, we introduce the weighted rank change (WRC) for each item and then aggregate them to obtain the system-level rank fluctuation on the daily basis. By grouping items into influx, outflux, and intra-flux, we reveal that the influx influences most the aggregate WRC, mostly due to the newly released items having got popular right away. For the individual-level analysis, we measure the lifetime, first rank, last rank, and highest rank of each item. We find that (i) the lifetime distributions are heavy-tailed, (ii) first rank distributions are more uniform than last rank distributions showing clear peaks at the bottom positions, (iii) ranks are overall positively correlated with each other, and (iv) the lifetimes tend to have the strongest positive correlation with the highest ranks. All these findings help us gain insights into the ranking dynamics of movies and music.


As for future works, we consider two directions. The first is to apply our system-level or group-level WRC measures to study the effects of shocks, whether being intrinsic or external, on the ranking dynamics. The second is to relate the ranking dynamics to the score dynamics. Ranks are simpler than scores, whereas scores contain more detailed information on the items' performance or competitiveness than ranks. Thus, selecting an appropriate weight function to reflect score disparities remains a promising direction for future research.

\section*{Competing interests}
  The authors have no competing interests.

\section*{Author's contributions}
H.-H.J., G.I., and H.J.P. designed the study and developed the research framework. H.-W.L. collected and analyzed the data. 
All authors discussed the results and contributed to writing and revising the manuscript.

\section*{Acknowledgements}
H.-H.J. acknowledges financial support by the National Research Foundation of Korea (NRF) grant funded by the Korea government (MSIT) (No. 2022R1A2C1007358).
H.J.P is supported by the National Research Foundation of Korea grant funded by the Korean government (MSIT) (Grant No. NRF-2022S1A5A2A03051182, RS-2023-00214071, and RS-2024-00460958). 

\bibliographystyle{apsrev4-1}

\begin{thebibliography}{30}%
\makeatletter
\providecommand \@ifxundefined [1]{%
 \@ifx{#1\undefined}
}%
\providecommand \@ifnum [1]{%
 \ifnum #1\expandafter \@firstoftwo
 \else \expandafter \@secondoftwo
 \fi
}%
\providecommand \@ifx [1]{%
 \ifx #1\expandafter \@firstoftwo
 \else \expandafter \@secondoftwo
 \fi
}%
\providecommand \natexlab [1]{#1}%
\providecommand \enquote  [1]{``#1''}%
\providecommand \bibnamefont  [1]{#1}%
\providecommand \bibfnamefont [1]{#1}%
\providecommand \citenamefont [1]{#1}%
\providecommand \href@noop [0]{\@secondoftwo}%
\providecommand \href [0]{\begingroup \@sanitize@url \@href}%
\providecommand \@href[1]{\@@startlink{#1}\@@href}%
\providecommand \@@href[1]{\endgroup#1\@@endlink}%
\providecommand \@sanitize@url [0]{\catcode `\\12\catcode `\$12\catcode
  `\&12\catcode `\#12\catcode `\^12\catcode `\_12\catcode `\%12\relax}%
\providecommand \@@startlink[1]{}%
\providecommand \@@endlink[0]{}%
\providecommand \url  [0]{\begingroup\@sanitize@url \@url }%
\providecommand \@url [1]{\endgroup\@href {#1}{\urlprefix }}%
\providecommand \urlprefix  [0]{URL }%
\providecommand \Eprint [0]{\href }%
\providecommand \doibase [0]{http://dx.doi.org/}%
\providecommand \selectlanguage [0]{\@gobble}%
\providecommand \bibinfo  [0]{\@secondoftwo}%
\providecommand \bibfield  [0]{\@secondoftwo}%
\providecommand \translation [1]{[#1]}%
\providecommand \BibitemOpen [0]{}%
\providecommand \bibitemStop [0]{}%
\providecommand \bibitemNoStop [0]{.\EOS\space}%
\providecommand \EOS [0]{\spacefactor3000\relax}%
\providecommand \BibitemShut  [1]{\csname bibitem#1\endcsname}%
\let\auto@bib@innerbib\@empty
\bibitem [{\citenamefont {Radicchi}\ \emph {et~al.}(2009)\citenamefont
  {Radicchi}, \citenamefont {Fortunato}, \citenamefont {Markines},\ and\
  \citenamefont {Vespignani}}]{Radicchi2009}%
  \BibitemOpen
  \bibfield  {author} {\bibinfo {author} {\bibfnamefont {F.}~\bibnamefont
  {Radicchi}}, \bibinfo {author} {\bibfnamefont {S.}~\bibnamefont {Fortunato}},
  \bibinfo {author} {\bibfnamefont {B.}~\bibnamefont {Markines}}, \ and\
  \bibinfo {author} {\bibfnamefont {A.}~\bibnamefont {Vespignani}},\ }\href
  {\doibase 10.1103/PhysRevE.80.056103} {\bibfield  {journal} {\bibinfo
  {journal} {Phys. Rev. E}\ }\textbf {\bibinfo {volume} {80}},\ \bibinfo
  {pages} {056103} (\bibinfo {year} {2009})}\BibitemShut {NoStop}%
\bibitem [{\citenamefont {Langville}\ and\ \citenamefont
  {Meyer}(2012)}]{Langville2012}%
  \BibitemOpen
  \bibfield  {author} {\bibinfo {author} {\bibfnamefont {A.~N.}\ \bibnamefont
  {Langville}}\ and\ \bibinfo {author} {\bibfnamefont {C.~D.}\ \bibnamefont
  {Meyer}},\ }\href@noop {} {\emph {\bibinfo {title} {Who’s \#1?: The Science
  of Rating and Ranking}}}\ (\bibinfo {address} {Princeton University Press},\
  \bibinfo {year} {2012})\BibitemShut {NoStop}%
\bibitem [{\citenamefont {Merritt}\ and\ \citenamefont
  {Clauset}(2014)}]{merritt_scoring_2014}%
  \BibitemOpen
  \bibfield  {author} {\bibinfo {author} {\bibfnamefont {S.}~\bibnamefont
  {Merritt}}\ and\ \bibinfo {author} {\bibfnamefont {A.}~\bibnamefont
  {Clauset}},\ }\href {\doibase 10.1140/epjds29} {\bibfield  {journal}
  {\bibinfo  {journal} {EPJ Data Science}\ }\textbf {\bibinfo {volume} {3}},\
  \bibinfo {pages} {4} (\bibinfo {year} {2014})}\BibitemShut {NoStop}%
\bibitem [{\citenamefont {Yucesoy}\ and\ \citenamefont
  {Barab{\'a}si}(2016)}]{yucesoy_untangling_2016}%
  \BibitemOpen
  \bibfield  {author} {\bibinfo {author} {\bibfnamefont {B.}~\bibnamefont
  {Yucesoy}}\ and\ \bibinfo {author} {\bibfnamefont {A.-L.}\ \bibnamefont
  {Barab{\'a}si}},\ }\href {\doibase 10.1140/epjds/s13688-016-0079-z}
  {\bibfield  {journal} {\bibinfo  {journal} {EPJ Data Science}\ }\textbf
  {\bibinfo {volume} {5}},\ \bibinfo {pages} {17} (\bibinfo {year}
  {2016})}\BibitemShut {NoStop}%
\bibitem [{\citenamefont {\'Erdi}(2019)}]{Erdi2019}%
  \BibitemOpen
  \bibfield  {author} {\bibinfo {author} {\bibfnamefont {P.}~\bibnamefont
  {\'Erdi}},\ }\href@noop {} {\emph {\bibinfo {title} {Ranking: The Unwritten
  Rules of The Social Game We All Play}}}\ (\bibinfo {address} {Oxford
  University Press},\ \bibinfo {year} {2019})\BibitemShut {NoStop}%
\bibitem [{\citenamefont {Zipf}(1949)}]{Zipf}%
  \BibitemOpen
  \bibfield  {author} {\bibinfo {author} {\bibfnamefont {G.~K.}\ \bibnamefont
  {Zipf}},\ }\href@noop {} {\emph {\bibinfo {title} {Human Behavior and the
  Principle of Least Effort: An Introduction to Human Ecology}}}\ (\bibinfo
  {publisher} {Addison-Wesley Press},\ \bibinfo {address} {Cambridge, MA,
  U.S.A.},\ \bibinfo {year} {1949})\BibitemShut {NoStop}%
\bibitem [{\citenamefont {Margalef}(1994)}]{margalef1994}%
  \BibitemOpen
  \bibfield  {author} {\bibinfo {author} {\bibfnamefont {R.}~\bibnamefont
  {Margalef}},\ }\href@noop {} {\bibfield  {journal} {\bibinfo  {journal} {Sci.
  Mar.}\ }\textbf {\bibinfo {volume} {58}},\ \bibinfo {pages} {87} (\bibinfo
  {year} {1994})}\BibitemShut {NoStop}%
\bibitem [{\citenamefont {Axtell}(2001)}]{Axtell2001}%
  \BibitemOpen
  \bibfield  {author} {\bibinfo {author} {\bibfnamefont {R.~L.}\ \bibnamefont
  {Axtell}},\ }\href@noop {} {\bibfield  {journal} {\bibinfo  {journal}
  {Science}\ }\textbf {\bibinfo {volume} {293}},\ \bibinfo {pages} {1818}
  (\bibinfo {year} {2001})}\BibitemShut {NoStop}%
\bibitem [{\citenamefont {Adamic}\ and\ \citenamefont
  {Huberman}(2002)}]{adamic2002zipf}%
  \BibitemOpen
  \bibfield  {author} {\bibinfo {author} {\bibfnamefont {L.~A.}\ \bibnamefont
  {Adamic}}\ and\ \bibinfo {author} {\bibfnamefont {B.~A.}\ \bibnamefont
  {Huberman}},\ }\href@noop {} {\bibfield  {journal} {\bibinfo  {journal}
  {Glottometrics}\ }\textbf {\bibinfo {volume} {3}},\ \bibinfo {pages} {143}
  (\bibinfo {year} {2002})}\BibitemShut {NoStop}%
\bibitem [{\citenamefont {Newman}(2005)}]{Newman01092005}%
  \BibitemOpen
  \bibfield  {author} {\bibinfo {author} {\bibfnamefont {M.}~\bibnamefont
  {Newman}},\ }\href {\doibase 10.1080/00107510500052444} {\bibfield  {journal}
  {\bibinfo  {journal} {Contemporary Physics}\ }\textbf {\bibinfo {volume}
  {46}},\ \bibinfo {pages} {323} (\bibinfo {year} {2005})},\ \Eprint
  {http://arxiv.org/abs/https://doi.org/10.1080/00107510500052444}
  {https://doi.org/10.1080/00107510500052444} \BibitemShut {NoStop}%
\bibitem [{\citenamefont {Pueyo}(2006)}]{Pueyo2006}%
  \BibitemOpen
  \bibfield  {author} {\bibinfo {author} {\bibfnamefont {S.}~\bibnamefont
  {Pueyo}},\ }\href {\doibase https://doi.org/10.1111/j.0030-1299.2006.14184.x}
  {\bibfield  {journal} {\bibinfo  {journal} {Oikos}\ }\textbf {\bibinfo
  {volume} {112}},\ \bibinfo {pages} {156} (\bibinfo {year} {2006})},\ \Eprint
  {http://arxiv.org/abs/https://nsojournals.onlinelibrary.wiley.com/doi/pdf/10.1111/j.0030-1299.2006.14184.x}
  {https://nsojournals.onlinelibrary.wiley.com/doi/pdf/10.1111/j.0030-1299.2006.14184.x}
  \BibitemShut {NoStop}%
\bibitem [{\citenamefont {Baek}\ \emph {et~al.}(2011)\citenamefont {Baek},
  \citenamefont {Bernhardsson},\ and\ \citenamefont {Minnhagen}}]{Baek_2011}%
  \BibitemOpen
  \bibfield  {author} {\bibinfo {author} {\bibfnamefont {S.~K.}\ \bibnamefont
  {Baek}}, \bibinfo {author} {\bibfnamefont {S.}~\bibnamefont {Bernhardsson}},
  \ and\ \bibinfo {author} {\bibfnamefont {P.}~\bibnamefont {Minnhagen}},\
  }\href {\doibase 10.1088/1367-2630/13/4/043004} {\bibfield  {journal}
  {\bibinfo  {journal} {New Journal of Physics}\ }\textbf {\bibinfo {volume}
  {13}},\ \bibinfo {pages} {043004} (\bibinfo {year} {2011})}\BibitemShut
  {NoStop}%
\bibitem [{\citenamefont {Deng}\ \emph {et~al.}(2012)\citenamefont {Deng},
  \citenamefont {Li}, \citenamefont {Cai}, \citenamefont {Bulou},\ and\
  \citenamefont {Wang}}]{Weibing2012universal}%
  \BibitemOpen
  \bibfield  {author} {\bibinfo {author} {\bibfnamefont {W.}~\bibnamefont
  {Deng}}, \bibinfo {author} {\bibfnamefont {W.}~\bibnamefont {Li}}, \bibinfo
  {author} {\bibfnamefont {X.}~\bibnamefont {Cai}}, \bibinfo {author}
  {\bibfnamefont {A.}~\bibnamefont {Bulou}}, \ and\ \bibinfo {author}
  {\bibfnamefont {Q.~A.}\ \bibnamefont {Wang}},\ }\href@noop {} {\bibfield
  {journal} {\bibinfo  {journal} {New Journal of Physics}\ }\textbf {\bibinfo
  {volume} {14}},\ \bibinfo {pages} {093038} (\bibinfo {year}
  {2012})}\BibitemShut {NoStop}%
\bibitem [{\citenamefont {Arshad}\ \emph {et~al.}(2018)\citenamefont {Arshad},
  \citenamefont {Hu},\ and\ \citenamefont {Ashraf}}]{ARSHAD201875}%
  \BibitemOpen
  \bibfield  {author} {\bibinfo {author} {\bibfnamefont {S.}~\bibnamefont
  {Arshad}}, \bibinfo {author} {\bibfnamefont {S.}~\bibnamefont {Hu}}, \ and\
  \bibinfo {author} {\bibfnamefont {B.~N.}\ \bibnamefont {Ashraf}},\ }\href
  {\doibase https://doi.org/10.1016/j.physa.2017.10.005} {\bibfield  {journal}
  {\bibinfo  {journal} {Physica A: Statistical Mechanics and its Applications}\
  }\textbf {\bibinfo {volume} {492}},\ \bibinfo {pages} {75} (\bibinfo {year}
  {2018})}\BibitemShut {NoStop}%
\bibitem [{\citenamefont {Houssard}\ \emph {et~al.}(2023)\citenamefont
  {Houssard}, \citenamefont {Pilati},\ and\ \citenamefont {et~al.}}]{twitchtv}%
  \BibitemOpen
  \bibfield  {author} {\bibinfo {author} {\bibfnamefont {A.}~\bibnamefont
  {Houssard}}, \bibinfo {author} {\bibfnamefont {M.}~\bibnamefont {Pilati},
  \bibfnamefont {F.and~Tartari}}, \ and\ \bibinfo {author} {\bibnamefont
  {et~al.}},\ }\href@noop {} {\bibfield  {journal} {\bibinfo  {journal}
  {Scientific Reports.}\ }\textbf {\bibinfo {volume} {13}},\ \bibinfo {pages}
  {1103} (\bibinfo {year} {2023})}\BibitemShut {NoStop}%
\bibitem [{\citenamefont {Barthelemy}(2023)}]{Barthelemy_2023}%
  \BibitemOpen
  \bibfield  {author} {\bibinfo {author} {\bibfnamefont {M.}~\bibnamefont
  {Barthelemy}},\ }\href {\doibase 10.1088/1361-6633/ace45e} {\bibfield
  {journal} {\bibinfo  {journal} {Reports on Progress in Physics}\ }\textbf
  {\bibinfo {volume} {86}},\ \bibinfo {pages} {084001} (\bibinfo {year}
  {2023})}\BibitemShut {NoStop}%
\bibitem [{\citenamefont {Cocho}\ \emph {et~al.}(2015)\citenamefont {Cocho},
  \citenamefont {Flores}, \citenamefont {Gershenson}, \citenamefont {Pineda},\
  and\ \citenamefont {S{\'a}nchez}}]{cocho2015rank}%
  \BibitemOpen
  \bibfield  {author} {\bibinfo {author} {\bibfnamefont {G.}~\bibnamefont
  {Cocho}}, \bibinfo {author} {\bibfnamefont {J.}~\bibnamefont {Flores}},
  \bibinfo {author} {\bibfnamefont {C.}~\bibnamefont {Gershenson}}, \bibinfo
  {author} {\bibfnamefont {C.}~\bibnamefont {Pineda}}, \ and\ \bibinfo {author}
  {\bibfnamefont {S.}~\bibnamefont {S{\'a}nchez}},\ }\href@noop {} {\bibfield
  {journal} {\bibinfo  {journal} {PLoS One}\ }\textbf {\bibinfo {volume}
  {10}},\ \bibinfo {pages} {e0121898} (\bibinfo {year} {2015})}\BibitemShut
  {NoStop}%
\bibitem [{\citenamefont {Morales}\ \emph
  {et~al.}(2018{\natexlab{a}})\citenamefont {Morales}, \citenamefont {Colman},
  \citenamefont {S{\'a}nchez}, \citenamefont {S{\'a}nchez-Puig}, \citenamefont
  {Pineda}, \citenamefont {I{\~n}iguez}, \citenamefont {Cocho}, \citenamefont
  {Flores},\ and\ \citenamefont {Gershenson}}]{morales2018rank}%
  \BibitemOpen
  \bibfield  {author} {\bibinfo {author} {\bibfnamefont {J.~A.}\ \bibnamefont
  {Morales}}, \bibinfo {author} {\bibfnamefont {E.}~\bibnamefont {Colman}},
  \bibinfo {author} {\bibfnamefont {S.}~\bibnamefont {S{\'a}nchez}}, \bibinfo
  {author} {\bibfnamefont {F.}~\bibnamefont {S{\'a}nchez-Puig}}, \bibinfo
  {author} {\bibfnamefont {C.}~\bibnamefont {Pineda}}, \bibinfo {author}
  {\bibfnamefont {G.}~\bibnamefont {I{\~n}iguez}}, \bibinfo {author}
  {\bibfnamefont {G.}~\bibnamefont {Cocho}}, \bibinfo {author} {\bibfnamefont
  {J.}~\bibnamefont {Flores}}, \ and\ \bibinfo {author} {\bibfnamefont
  {C.}~\bibnamefont {Gershenson}},\ }\href@noop {} {\bibfield  {journal}
  {\bibinfo  {journal} {Frontiers in Physics}\ }\textbf {\bibinfo {volume}
  {6}},\ \bibinfo {pages} {45} (\bibinfo {year}
  {2018}{\natexlab{a}})}\BibitemShut {NoStop}%
\bibitem [{\citenamefont {Blumm}\ \emph {et~al.}(2012)\citenamefont {Blumm},
  \citenamefont {Ghoshal}, \citenamefont {Forr\'o}, \citenamefont {Schich},
  \citenamefont {Bianconi}, \citenamefont {Bouchaud},\ and\ \citenamefont
  {Barab\'asi}}]{Blumm2012}%
  \BibitemOpen
  \bibfield  {author} {\bibinfo {author} {\bibfnamefont {N.}~\bibnamefont
  {Blumm}}, \bibinfo {author} {\bibfnamefont {G.}~\bibnamefont {Ghoshal}},
  \bibinfo {author} {\bibfnamefont {Z.}~\bibnamefont {Forr\'o}}, \bibinfo
  {author} {\bibfnamefont {M.}~\bibnamefont {Schich}}, \bibinfo {author}
  {\bibfnamefont {G.}~\bibnamefont {Bianconi}}, \bibinfo {author}
  {\bibfnamefont {J.-P.}\ \bibnamefont {Bouchaud}}, \ and\ \bibinfo {author}
  {\bibfnamefont {A.-L.}\ \bibnamefont {Barab\'asi}},\ }\href {\doibase
  10.1103/PhysRevLett.109.128701} {\bibfield  {journal} {\bibinfo  {journal}
  {Phys. Rev. Lett.}\ }\textbf {\bibinfo {volume} {109}},\ \bibinfo {pages}
  {128701} (\bibinfo {year} {2012})}\BibitemShut {NoStop}%
\bibitem [{\citenamefont {Morales}\ \emph {et~al.}(2016)\citenamefont
  {Morales}, \citenamefont {Sánchez}, \citenamefont {Flores},\ and\
  \citenamefont {et~al.}}]{temporal}%
  \BibitemOpen
  \bibfield  {author} {\bibinfo {author} {\bibfnamefont {J.}~\bibnamefont
  {Morales}}, \bibinfo {author} {\bibfnamefont {S.}~\bibnamefont {Sánchez}},
  \bibinfo {author} {\bibfnamefont {J.}~\bibnamefont {Flores}}, \ and\ \bibinfo
  {author} {\bibnamefont {et~al.}},\ }\href@noop {} {\bibfield  {journal}
  {\bibinfo  {journal} {EPJ Data Sci.}\ }\textbf {\bibinfo {volume} {5}},\
  \bibinfo {pages} {33} (\bibinfo {year} {2016})}\BibitemShut {NoStop}%
\bibitem [{\citenamefont {Morales}\ \emph
  {et~al.}(2018{\natexlab{b}})\citenamefont {Morales}, \citenamefont {Colman},
  \citenamefont {S\'anchez}, \citenamefont {S\'{a}nchez-Puig}, \citenamefont
  {Pineda}, \citenamefont {I\~{n}iguez}, \citenamefont {Cocho}, \citenamefont
  {Flores},\ and\ \citenamefont {Gershenson}}]{Morales2018}%
  \BibitemOpen
  \bibfield  {author} {\bibinfo {author} {\bibfnamefont {J.}~\bibnamefont
  {Morales}}, \bibinfo {author} {\bibfnamefont {E.}~\bibnamefont {Colman}},
  \bibinfo {author} {\bibfnamefont {S.}~\bibnamefont {S\'anchez}}, \bibinfo
  {author} {\bibfnamefont {F.}~\bibnamefont {S\'{a}nchez-Puig}}, \bibinfo
  {author} {\bibfnamefont {C.}~\bibnamefont {Pineda}}, \bibinfo {author}
  {\bibfnamefont {G.}~\bibnamefont {I\~{n}iguez}}, \bibinfo {author}
  {\bibfnamefont {G.}~\bibnamefont {Cocho}}, \bibinfo {author} {\bibfnamefont
  {J.}~\bibnamefont {Flores}}, \ and\ \bibinfo {author} {\bibfnamefont
  {C.}~\bibnamefont {Gershenson}},\ }\href {\doibase 10.3389/fphy.2018.00045}
  {\bibfield  {journal} {\bibinfo  {journal} {Frontiers in Physics}\ }\textbf
  {\bibinfo {volume} {6}} (\bibinfo {year} {2018}{\natexlab{b}}),\
  10.3389/fphy.2018.00045}\BibitemShut {NoStop}%
\bibitem [{\citenamefont {I\~{n}iguez}\ \emph {et~al.}(2022)\citenamefont
  {I\~{n}iguez}, \citenamefont {Pineda}, \citenamefont {Gershenson},\ and\
  \citenamefont {Barab\'{a}si}}]{iniguez2022}%
  \BibitemOpen
  \bibfield  {author} {\bibinfo {author} {\bibfnamefont {G.}~\bibnamefont
  {I\~{n}iguez}}, \bibinfo {author} {\bibfnamefont {C.}~\bibnamefont {Pineda}},
  \bibinfo {author} {\bibfnamefont {C.}~\bibnamefont {Gershenson}}, \ and\
  \bibinfo {author} {\bibfnamefont {A.-L.}\ \bibnamefont {Barab\'{a}si}},\
  }\href@noop {} {\bibfield  {journal} {\bibinfo  {journal} {Nat. Commun.}\
  }\textbf {\bibinfo {volume} {13}},\ \bibinfo {pages} {1646} (\bibinfo {year}
  {2022})}\BibitemShut {NoStop}%
\bibitem [{Fli()}]{FlixPatrol}%
  \BibitemOpen
  \href@noop {} {\enquote {\bibinfo {title} {{{FlixPatrol}}},}\ }\bibinfo
  {howpublished} {https://flixpatrol.com}\BibitemShut {NoStop}%
\bibitem [{Kag()}]{Kaggle}%
  \BibitemOpen
  \href
  {https://www.kaggle.com/datasets/edumucelli/spotifys-worldwide-daily-song-ranking}
  {\enquote {\bibinfo {title} {Spotify music data},}\ }\BibitemShut {NoStop}%
\bibitem [{\citenamefont {Murase}\ \emph {et~al.}(2019)\citenamefont {Murase},
  \citenamefont {Jo}, \citenamefont {T{\"o}r{\"o}k}, \citenamefont
  {Kert{\'e}sz},\ and\ \citenamefont {Kaski}}]{Murase2019Sampling}%
  \BibitemOpen
  \bibfield  {author} {\bibinfo {author} {\bibfnamefont {Y.}~\bibnamefont
  {Murase}}, \bibinfo {author} {\bibfnamefont {H.-H.}\ \bibnamefont {Jo}},
  \bibinfo {author} {\bibfnamefont {J.}~\bibnamefont {T{\"o}r{\"o}k}}, \bibinfo
  {author} {\bibfnamefont {J.}~\bibnamefont {Kert{\'e}sz}}, \ and\ \bibinfo
  {author} {\bibfnamefont {K.}~\bibnamefont {Kaski}},\ }\href {\doibase
  10.1103/PhysRevE.99.052304} {\bibfield  {journal} {\bibinfo  {journal}
  {Physical Review E}\ }\textbf {\bibinfo {volume} {99}},\ \bibinfo {pages}
  {052304} (\bibinfo {year} {2019})}\BibitemShut {NoStop}%
\bibitem [{\citenamefont {Stoica}\ and\ \citenamefont
  {Moses}(2005)}]{stoica2005spectral}%
  \BibitemOpen
  \bibfield  {author} {\bibinfo {author} {\bibfnamefont {P.}~\bibnamefont
  {Stoica}}\ and\ \bibinfo {author} {\bibfnamefont {R.}~\bibnamefont {Moses}},\
  }\href@noop {} {\emph {\bibinfo {title} {Spectral Analysis of Signals}}}\
  (\bibinfo  {publisher} {Pearson Prentice Hall},\ \bibinfo {year}
  {2005})\BibitemShut {NoStop}%
\bibitem [{\citenamefont {Es}\ and\ \citenamefont {Nguyen}(2025)}]{MaC9362}%
  \BibitemOpen
  \bibfield  {author} {\bibinfo {author} {\bibfnamefont {K.}~\bibnamefont
  {Es}}\ and\ \bibinfo {author} {\bibfnamefont {D.}~\bibnamefont {Nguyen}},\
  }\href {\doibase 10.17645/mac.9362} {\bibfield  {journal} {\bibinfo
  {journal} {Media and Communication}\ }\textbf {\bibinfo {volume} {13}}
  (\bibinfo {year} {2025}),\ 10.17645/mac.9362}\BibitemShut {NoStop}%
\bibitem [{\citenamefont {Zhang}\ \emph {et~al.}(2013)\citenamefont {Zhang},
  \citenamefont {Kreitz}, \citenamefont {Isaksson}, \citenamefont {Ubillos},
  \citenamefont {Urdaneta}, \citenamefont {Pouwelse},\ and\ \citenamefont
  {Epema}}]{6566767}%
  \BibitemOpen
  \bibfield  {author} {\bibinfo {author} {\bibfnamefont {B.}~\bibnamefont
  {Zhang}}, \bibinfo {author} {\bibfnamefont {G.}~\bibnamefont {Kreitz}},
  \bibinfo {author} {\bibfnamefont {M.}~\bibnamefont {Isaksson}}, \bibinfo
  {author} {\bibfnamefont {J.}~\bibnamefont {Ubillos}}, \bibinfo {author}
  {\bibfnamefont {G.}~\bibnamefont {Urdaneta}}, \bibinfo {author}
  {\bibfnamefont {J.~A.}\ \bibnamefont {Pouwelse}}, \ and\ \bibinfo {author}
  {\bibfnamefont {D.}~\bibnamefont {Epema}},\ }in\ \href {\doibase
  10.1109/INFCOM.2013.6566767} {\emph {\bibinfo {booktitle} {2013 Proceedings
  IEEE INFOCOM}}}\ (\bibinfo {year} {2013})\ pp.\ \bibinfo {pages}
  {220--224}\BibitemShut {NoStop}%
\bibitem [{\citenamefont {Wu}\ and\ \citenamefont {Huberman}(2007)}]{Wu2007}%
  \BibitemOpen
  \bibfield  {author} {\bibinfo {author} {\bibfnamefont {F.}~\bibnamefont
  {Wu}}\ and\ \bibinfo {author} {\bibfnamefont {B.~A.}\ \bibnamefont
  {Huberman}},\ }\href@noop {} {\bibfield  {journal} {\bibinfo  {journal}
  {Proceedings of the National Academy of Sciences}\ }\textbf {\bibinfo
  {volume} {104(45)}},\ \bibinfo {pages} {17599} (\bibinfo {year}
  {2007})}\BibitemShut {NoStop}%
\bibitem [{\citenamefont {Crane}\ and\ \citenamefont
  {Sornette}(2008)}]{crane2008}%
  \BibitemOpen
  \bibfield  {author} {\bibinfo {author} {\bibfnamefont {R.}~\bibnamefont
  {Crane}}\ and\ \bibinfo {author} {\bibfnamefont {D.}~\bibnamefont
  {Sornette}},\ }\href@noop {} {\bibfield  {journal} {\bibinfo  {journal}
  {Proceedings of the National Academy of Sciences}\ }\textbf {\bibinfo
  {volume} {105(41)}},\ \bibinfo {pages} {15649} (\bibinfo {year}
  {2008})}\BibitemShut {NoStop}%
\end{thebibliography}
%

\end{document}